\documentclass[%
reprint, 
amsmath,amssymb,
aps,showkeys
 ]{revtex4-2}

\usepackage{graphicx}
\usepackage{dcolumn}
\usepackage{bm}
\usepackage{hyperref}
\usepackage[mathlines]{lineno}
\usepackage{amssymb}

\begin{document}
	
	\title{Schwinger's SUSY Oscillators: An Analysis}
	
	\author{Dheeraj Shukla$^1$}
	\email{shudheer.phy@gmail.com}
	 \author{Sudhaker Upadhyay$^{2,3,4}$\footnote{Corresponding Author}\footnote{Visiting Associate, Inter-University Centre for Astronomy and Astrophysics (IUCAA) Pune-411007, Maharashtra, India}}
	 \email{sudhakerupadhyay@gmail.com}
	\affiliation{$^1$Department of Physics, Panjab University, Chandigarh-160014, India}	
	
	 \affiliation{$^2$Department of Physics, K.L.S. College, Nawwada, Magadh University, Bodh Gaya,  Bihar 805110,  India}
	 	 \affiliation{$^3$School of Physics, Damghan University, Damghan, 3671641167, Iran} 
	 	 	 \affiliation{$^4$Canadian Quantum Research Center 204-3002 32 Ave Vernon, BC V1T 2L7 Canada}

	\date{\today}
	
	\begin{abstract}
	In this article, we explore the inconsistencies in the physics of fermionic oscillators and propose potential solutions to address them. By rigorously deriving the Hamiltonian and Lagrangian from first principles, we aim to provide a comprehensive and fundamental understanding of the system. Furthermore, we calculate the partition function for a system of fermionic oscillators by drawing a direct analogy to Planck's treatment of energy distribution in bosonic oscillators, offering a parallel approach to this well-established method.	
	 Our study extends beyond the conventional framework by investigating the generalized angular momentum algebra within the context of Schwinger's oscillator model. This includes a detailed examination of the algebraic structures for combinations of bosonic-bosonic, bosonic-fermionic, and fermionic-fermionic oscillators. Through this, we delve into these systems' underlying symmetries and algebraic richness, shedding light on the intricate relationships between these different types of oscillators.
	 In addition to these foundational aspects, we explore the broader implications of this generalized Schwinger approach. Our analysis touches upon potential applications and consequences of this formalism, offering insights that could be relevant to various areas of theoretical physics. This work paves the way for a deeper understanding of quantum oscillators and their role in modern physics by bridging the gap between bosonic and fermionic oscillators.
	\end{abstract}
	
	\maketitle

\section{Introduction}
The quantum harmonic oscillator is a cornerstone in quantum mechanics, providing a solvable model approximating many physical systems near their equilibrium points. Apart from this, it plays a crucial role in quantum field theory and information science, aiding in developing quantum simulations and computing paradigms \cite{Woit}. Further its applications span various fields such as superconductivity \cite{Fonzin,Yang}, exotic material science \cite{Devoret, Matheny, Akash}, quantum informatics, nuclear physics, astrophysics, and quantum many-body systems.

As the fermionic counterpart to the bosonic harmonic oscillator, the fermionic oscillator presents unique theoretical challenges. Unlike bosonic oscillators, fermionic oscillators lack direct physical realizations, leading to a theoretical framework built on analogies to bosonic systems. The Hamiltonian for fermionic oscillators is often introduced using classical conjugate Grassmann variables, which obey specific anticommutation relations. These variables exist in Grassmann space, distinct from the spatial coordinates and velocities used in bosonic systems.

In quantum mechanics, the treatment of fermionic oscillators diverges from the bosonic case. While bosonic oscillators use creation and annihilation operators ($a$) and ($a^\dagger$), fermionic oscillators employ operators ($f$) and ($f^\dagger$) that satisfy anticommutation relations. The path-integral approach is often used to explore the statistical properties of fermionic oscillators, leading to the formulation of a partition function for Grassmann fields.

Schwinger’s approach, as discussed by Sakurai \cite{Sakurai}, shows that the algebra of a pair of independent oscillators is equivalent to the angular momentum algebra of a rotor \cite{Woit}. This indicates deeper underlying relationships and suggests an inherent supersymmetry, which can be applied to study collections of fermions oscillating at various frequencies and species \cite{Woit}.

The paper is presented as follows. In section \ref{sec2}, we discuss the fermionic oscillator following the bosonic oscillator. The Hamiltonian for the fermionic oscillator is calculated from the first principle in section \ref{sec3}.
In particular, the path integral approach is an alternate way to calculate the fermionic oscillator in section \ref{sec4}. The time period for the fermionic oscillator is calculated in section \ref{sec5}. We discuss the fermionic algebra representation in section \ref{sec6}. Schwinger's algebra of a system of two simultaneous harmonic oscillators is addressed in section \ref{sec7}. The physical realization of Schwinger's harmonic oscillators is studied in section \ref{sec8}. We added a discussion on the fermionic oscillator and Schwinger formalism in section \ref{sec9}. We summarize the work with future remarks in the last section.

\section{Fermionic Oscillator}\label{sec2}
A fermionic oscillator is the fermionic version of a bosonic harmonic oscillator. Unfortunately, we do not have any fermionic oscillators as we have in the case of bosonic oscillators, e.g. simple and torsional pendulums, spring oscillations, etc. In most of the literature, this problem has been studied in analogy with bosonic oscillators. We lack a derivation of its Hamiltonian or Lagrangian from the first principles we have in the case of bosonic oscillators. The reason for this insight might be hidden in the fact that we do not have any realization of such oscillators in tabletop experiments. 
	
The classical 1D Hamiltonian for an oscillator of mass $m$ oscillates with a frequency $\omega$ in terms of position $x$ and momentum $p$ as
\begin{eqnarray}
		H = \dfrac{p^2}{2m} + \dfrac{m}{2}~\omega^2 x^2(t).\label{eqn 1}
\end{eqnarray}
	
The fermionic classical counterpart for the Hamiltonian is  taken in an ad-hoc manner as follows
\begin{eqnarray}
		H = \dfrac{\hbar\omega}{2}~[\bar\psi,~\psi]\label{eqn 2},
\end{eqnarray}
where $\psi(t)/\bar\psi (t)$ are classical conjugate Grassmann variables subjected to the conditions
\begin{eqnarray}
		\psi\bar\psi + \bar\psi \psi = 0,~~\psi^2 = 0 = \bar\psi^2.\label{eqn 3}
\end{eqnarray}
	
One thing worth noting is that there appears to be no mass term in the fermionic Hamiltonians contrary to the bosonic Hamiltonian. Also, the fermionic variables are not related to location and velocity in spatial space. These exist in a very different kind of conjugate space called Grassmann space. 
	
The quantum Hamiltonian version for the bosonic oscillator in terms of operators $a$ and $a^\dagger$ is given as follows: 
\begin{eqnarray}
		H = \hbar\omega \left(a^\dagger a + \dfrac{1}{2}\right).\label{eqn 4}
\end{eqnarray}
	
The explicit forms of the operators $a$ and $a^\dagger$ are given as
\begin{eqnarray}
		a &&= \sqrt{\dfrac{\hbar}{2m\omega}}(x+ip) = \sqrt{\dfrac{\hbar}{2m\omega}}\left(x+\dfrac{d}{dx}\right),\nonumber\\
		a^\dagger && =\sqrt{\dfrac{\hbar}{2m\omega}}(x-ip) = \sqrt{\dfrac{\hbar}{2m\omega}}\left(x-\dfrac{d}{dx}\right), \label{eqn 5}
\end{eqnarray}
subjected to the condition $[a, a^\dagger] = 1$.
	
The quantum counterpart of this Hamiltonian for fermionic oscillators is taken to be in a logical but ad-hoc manner as follows:
\begin{eqnarray}
		H = \hbar\omega \left(f^\dagger f - \dfrac{1}{2}\right), \label{eqn 6}
\end{eqnarray}
subjected to the condition $\lbrace f, f^\dagger\rbrace = 1$.
	
A statistical cum path-integral approach to this oscillator problem is obtained in terms of the partition function for the Grassmann fields, defined as follows
\begin{eqnarray}
		Z =  \text{Tr}~e^{-\beta H}, \label{eqn 7}
\end{eqnarray}
where $\beta = 1/kT$ and $H$ is the Hamiltonian for the fermionic system. The partition function for the Grassmann fields $\psi (t)$ and $\bar\psi (t)$ after some calculations can be written as 
\begin{eqnarray}
		Z [0] = \int {\cal D}\psi\,{\cal D}\bar\psi~ e^{i{\cal A}(\psi, \bar\psi)/\hbar}, \label{eqn 8}
\end{eqnarray}
whereas the action for the fermionic oscillator is given as
\begin{eqnarray}
		{\cal A}~[\psi, \bar\psi] = \int dt \, \left(\dfrac{i\hbar}{2}\,(\bar\psi \dot \psi - \dot{\bar\psi} \psi) - \dfrac{\hbar\omega}{2}\,[\bar\psi, \psi]\right). \label{eqn 9}
\end{eqnarray}
	
Thus, the Lagrangian for fermionic oscillators reduces to the form
\begin{eqnarray}
		L = \dfrac{i\hbar}{2}\,(\bar\psi \dot \psi - \dot{\bar\psi} \psi) - \dfrac{\hbar\omega}{2}\,[\bar\psi, \psi], \label{eqn 10}
\end{eqnarray}
where the frequency of oscillator is $\omega$ and mass is $m$.
	
\section{Oscillator physics from the first principles}\label{sec3}

The Hamiltonian for the fermionic oscillator is obtained from the Dirac Lagrangian for the fermions in terms of the Dirac wave functions, which are Grassmann odd variables, i.e.  $\lbrace \bar\psi, \psi \rbrace = 0$. The Dirac Lagrangian is given as follows	
\begin{eqnarray}
		{\cal L} = \bar\psi(i\hbar\gamma^\mu\partial_\mu - mc)\psi, \label{eqn 11}
\end{eqnarray}
where $\gamma^\mu$ are Dirac matrices and are subject to the Clifford algebra, i.e. 
$\lbrace \gamma^\mu, \gamma^\nu\rbrace = 2\eta^{\mu\nu} I$ and the Minkowskian metric in matrix form is taken to be: $\eta_{\mu\nu} = \eta^{\mu\nu} = $ diag$(1, -1, -1, -1)$. 
	
For the fields $\psi(t)$ and $\bar\psi(t)$, the action for a fixed volume reduces to the form in terms of the Lagrangian, i.e. 
	
\begin{eqnarray}
		{\cal A}~ && = \int c~dt~ {\cal L}, \nonumber\\
		&& = \int c~dt~\bar\psi (i\hbar~\gamma^0 \partial_0 - mc)\psi, \nonumber\\
		&& = \int dt~ \left(i\hbar~\bar\psi \dot\psi - mc^2~ \bar\psi \psi \right),\nonumber\\
		&& = \int dt~\left(\dfrac{i\hbar}{2}~(\bar\psi \dot\psi - \dot{\bar\psi} \psi) - \dfrac{mc^2}{2}~[\bar\psi, \psi]\right), \label{eqn 12}
\end{eqnarray}	
where we have used the fact that $\gamma^ 0 = I$ for 1-dimensional case.
	
This reduced form of the action is same as the action proposed in an ad-hoc manner in (\ref{eqn 9}), provided we take $\hbar\omega = mc^2$. The corresponding momenta for $\psi$ and $\bar\psi$ are obtained as follows
\begin{eqnarray}
		&& \prod_\psi = \dfrac{\partial L}{\partial \dot \psi} = -\dfrac{i\hbar}{2}\bar\psi,\nonumber\\
		&& \prod_{\bar\psi} = \dfrac{\partial L}{\partial \dot{\bar\psi}} = -\dfrac{i\hbar}{2}\psi. \label{eqn 13}
	\end{eqnarray}
	The Hamiltonian for the above system can be obtained using the Legend\'re transformations and thus turns out to be \cite{Ashok, Woit} 
	\begin{eqnarray}
		H = \dfrac{\hbar\omega}{2}[\bar\psi, \psi]. \label{eqn 14} 
	\end{eqnarray}
	\section{An Alternative Way}	\label{sec4}
	
	This is a standard practice to evaluate the above using the path integral approach. However, we will calculate it similarly: the partition function for a bosonic oscillator is calculated. 
	
	The Hamiltonian for a fermionic oscillator in terms of the Dirac creation and annihilation operators, i.e. $f^\dagger$ and $f$, can be expressed as give in the references \cite{Ashok, Woit,Mejia}, i.e.,
	\begin{eqnarray}
		H = \hbar\omega \left(f^\dagger f - \dfrac{1}{2}\right). \label{eqn 15}
	\end{eqnarray}
		
	It is to be noted that this form of Hamiltonian is not derived through any first principle; instead, it has been taken in an ad-hoc manner in analogy with the quantum Hamiltonian for the bosonic oscillator. 
	
	It can be shown that a fermionic oscillator can have two basis states, namely, $\vert 0\rangle$ and $\vert 1\rangle$. The action of Hamiltonian for these two states is
	\begin{eqnarray}
	H\vert 0  \rangle = \dfrac{\hbar\omega}{2}\,\vert 0  \rangle; \qquad
	H\vert 1  \rangle = -\dfrac{\hbar\omega}{2}\,\vert 1  \rangle. \label{eqn 16}
	\end{eqnarray}
	
	Thus, the partition function for a single fermionic oscillator at a given temperature characterized by $\beta^{-1} = kT$ can be obtained as
	\begin{eqnarray}
	Z[0] && = \text{Tr} \,e^{\beta H} = \langle 0 \vert e^{\beta H}\vert 0\rangle + \langle 1 \vert e^{\beta H}\vert 1\rangle, \nonumber\\
	\Rightarrow Z[0] && = 2\cosh \dfrac{\beta\hbar\omega}{2}. \label{eqn 17}
	\end{eqnarray}
	
	Now, let us compute the $Z[0]$ for a collection of oscillators oscillating with frequencies $\omega_n$ for $n = 1,2,\cdots$. The resultant partition function is a product of the partition functions of the individual oscillators, i.e.
	\begin{eqnarray}
	Z[0] &&= Z_1 [0] \times Z_2[0] \times Z_3[0]\times \cdots \times Z_n[0], \nonumber\\
	&& = 2^n \,\cosh \dfrac{\beta\hbar\omega_1}{2} \,\cosh \dfrac{\beta\hbar\omega_2}{2}\,\cdots \cosh \dfrac{\beta\hbar\omega_n}{2}. \label{eqn 18}
	\end{eqnarray}

	The logarithm of the partition function is very useful in the calculation of statistical/thermodynamic quantities; therefore,
	\begin{eqnarray}
		\ln Z & =& n \ln 2 + \ln \cosh \dfrac{\hbar\beta\omega_1}{2} + \cdots \nonumber\\
		&+& \ln \cosh \dfrac{\hbar\beta\omega_n}{2}. \label{eqn 19}
	\end{eqnarray}
	
	This can be further simplified as
	\begin{eqnarray}
		\ln Z = \sum_n \left[\dfrac{\hbar\beta\omega_n}{2} + \ln \left(1 + e^{-\beta\hbar\omega_n}\right)\right]. \label{eqn 20}
	\end{eqnarray}	
	Therefore, we can have the total thermal energy of the system as 
	\begin{eqnarray}
		\langle E\rangle &&= - \dfrac{\partial \ln Z}{\partial\beta}, \nonumber\\ 
		&& = -\sum_n \dfrac{\hbar\omega_n}{2}\,\tanh\left(\dfrac{\beta\hbar\omega_n}{2}\right),\nonumber\\
		&& = -\sum_n \hbar \omega_n \left(\dfrac{1}{2} - \dfrac{1}{1+ e^{\beta\hbar\omega_n}}\right). \label{eqn 21}
	\end{eqnarray}	
	For a large collection of fermionic oscillators, i.e. in the continuum limit, we can compute the average energy to be 
	\begin{eqnarray}
		 \beta\langle E \rangle = \int_1^{\epsilon_m} dx ~ \left( \dfrac{1}{1+ e^x}
		 -\dfrac{1}{2} \right). \label{eqn 22}
	\end{eqnarray}
		Now, taking $x= \beta\hbar\omega$, the average energy can be written in terms of the upper limit $\epsilon_m =\beta \hbar\omega_m$ as \cite{Kaputsa, Laine} 
	\begin{eqnarray}
		\langle E \rangle = k T\, \left(\dfrac{1-\epsilon_m}{2} + \ln \dfrac{1+ e}{1+ e^{\epsilon_m}}\right), \label{eqn 23}
	\end{eqnarray}
at some temperature $T$ characterized by $\beta = 1/kT$ 
(where $k = 1.38\times 10^{-23} ~J/K$ is Boltzmann constant).
	
	It is interesting to compute the various thermodynamic potentials for this collection of fermionic oscillators in terms of the partition function $Z$. 
	
	This approach is far more compatible and straightforward than the one used in path integral formulation. However, this is not the last approach we discuss here.  Schwinger took a very n\"aive approach to the harmonic oscillator problem \cite{Sakurai}. Although he just did it for a straightforward version, the consequences seem far more promising, even capable enough to deal with the supersymmetric oscillators. 
	

	\section{Time periods of oscillations}\label{sec5}
	A 1-dimension equation for a bosonic oscillator is given as
	\[\ddot x + \omega^2 x = 0.\]
	The solution is a parametric curve in (1: 2) superspace in the form. 
	\[x_1(t) \sim e^{-i\omega t}, ~ x_2(t) \sim e^{i\omega t}. \label{eqn 24}\]
	
	The period $T$ of the oscillations can be obtained by the periodicity condition $x(t) = x(t + T)$ as $T = 2\pi/\omega $.
	
	The equation of oscillation for the fermionic oscillator can be obtained from the Euler-Lagrange equation of motion using right to left derivative on the Lagrangian (\ref{eqn 11})
	\begin{eqnarray}
		&& \dfrac{d}{dt}\left(\dfrac{\partial {\cal L}}{\overleftarrow{\partial}\dot\psi} \right) = \dfrac{\partial {\cal L}}{\overleftarrow{\partial} \psi},\nonumber\\
		\Rightarrow &&~\dot{\bar\psi}  = 2i\tilde \omega \bar\psi. \label{eqn 25}
	\end{eqnarray}
	
	Similarly, we can obtain the conjugate equation as 
	\[\dot \psi = -2i \tilde \omega \psi.\]
	The solutions of these equations are parametric curves in (1:2) superspace as
	\begin{eqnarray}
	\psi (t) \sim e^{-2i\tilde \omega t}, ~ \bar\psi (t) \sim e^{2i\tilde\omega t}. \label{eqn 26}
	\end{eqnarray}
	The period of oscillations can be analogously obtained from the periodicity conditions.
	\[\psi (t) = \psi(t+ T), ~ \bar\psi (t) = \bar\psi (t+ T). \label{eqn 27}\]
		This surprisingly gives the frequency of fermionic oscillations $\tilde \omega = \pi/T$. This is just half of the frequency of a bosonic oscillator oscillating with the same time period, i.e. $\tilde \omega = \omega/2$.

		
	\section{Representation of Fermionic Oscillator Algebra}\label{sec6}
	The simplest but non-trivial interpretation of the Dirac annihilation and creation operator is 
	\begin{eqnarray}
		f\equiv \theta, \quad f^\dagger \equiv \partial_\theta, \label{eqn 28}
	\end{eqnarray}
	with condition $\tilde N^2 = \tilde N$.		
	
	Apart from the above physical representation of annihilation and creation operators, one can realize them in terms of Pauli operators as evident here \cite{Woit}
	\begin{eqnarray}
		f = \dfrac{\sigma_1 - i \sigma_2}{2},\quad f^\dagger = \dfrac{\sigma_1 + i\sigma_2}{2},\quad H = \dfrac{\hbar\omega}{2}\sigma_3, \label{eqn 29}
	\end{eqnarray}
	which in turn is an irreducible representation that can be written as
	\begin{eqnarray}
		f = \begin{pmatrix}
			0 & 0\\
			1 & 0	
		\end{pmatrix},~ 
		f^\dagger = \begin{pmatrix}
			0 & 1\\
			0 & 0	
		\end{pmatrix}, ~
		\tilde N =  \begin{pmatrix}
			1 & 0\\
			0 & 0	
		\end{pmatrix}, \label{eqn 30}
	\end{eqnarray}
	on the fermionic states as
	\begin{eqnarray}
		\vert 0 \rangle = \begin{pmatrix}
			1\\
			0	
		\end{pmatrix}, ~ 
		\vert 1 \rangle = \begin{pmatrix}
			0\\
			1	
		\end{pmatrix}. \label{eqn 31}
	\end{eqnarray}
		If we express the fermionic fields in terms of some arbitrary Grassmann odd variables, say $\theta$ and $\bar\theta$, such as
	\begin{eqnarray}
		\psi = \dfrac{1}{\sqrt{2}}
		\begin{pmatrix} 
			\theta\\
			\bar\theta
		\end{pmatrix}, \quad 
		\bar\psi = \dfrac{1}{\sqrt{2}}
		\begin{pmatrix} 
			\bar\theta\\
			\theta
		\end{pmatrix}. \label{eqn 32}
	\end{eqnarray}
	
	The corresponding Lagrangian can be obtained as 
	\begin{eqnarray}
		L = \dfrac{i\hbar}{2} (\bar\theta \dot\theta - \dot{\bar\theta} \theta) - \dfrac{\hbar\omega}{2}(\bar\theta\theta - \theta\bar\theta), \label{eqn 33}
	\end{eqnarray}
	where we have taken $\hbar\omega = mc^2$ for the oscillator. This Lagrangian matches with the one obtained in (\ref{eqn 10}). It is also clear that the form of Lagrangian is independent of the fields, i.e. any two conjugate Grassmann odd variables/fields would yield this Lagrangian uniquely. 
	
	It is worth noting here that any set of conjugate Grassmann odd variables classically follows the anticommutation rule, but when treated quantum mechanically, the anticommutator turns into an identity, i.e.,
	\begin{eqnarray}
		&&\lbrace \theta, \bar\theta\rbrace = 0; \quad \text{classically}\nonumber\\
		&& \lbrace \theta, \bar\theta\rbrace = 1; \quad \text{quantum mechanically}. \label{eqn 34}
	\end{eqnarray}
		This is the essential criterion for dealing with a quantum fermionic or quantum fermionic system.


	\section{Schwinger's Harmonic Oscillators}\label{sec7}
	Julian Schwinger showed \cite{Sakurai} that the algebra of a system of two simultaneous harmonic oscillators is equivalent to an angular momentum algebra. Let us consider two harmonic oscillators with creation operators ($a_1^\dagger, a_2^\dagger $) and annihilation operators ($a_1, a_2 $) when taken together, are algebraically equivalent to a single particle orbiting around a common centre. 
	
	To show this, let us define the number operators with the help of the two harmonic oscillators as follows
	\begin{eqnarray}
		N_1 = a^\dagger_1  a_1, \quad N_2 = a_2^\dagger a_2.\label{35}
	\end{eqnarray}
		The harmonic oscillators algebra for $i =1,2$ can be written as
	\begin{eqnarray}
		[ a_i, a_i^\dagger] = 1, \; [N_i, a_i] = - a_i,\, [N_i, a^\dagger_i]= a_i^\dagger. \label{eqn 36}
	\end{eqnarray}
	
	The combined eigenstate of the two oscillators is given as 
	$\vert n_1, n_2 \rangle = \vert n_1 \rangle \otimes \vert n_2 \rangle$ such that\\
	$N_1\vert n_1, n_2 \rangle = n_1\vert n_1, n_2 \rangle$ and $N_2\vert n_1, n_2 \rangle = n_2\vert n_1, n_2 \rangle$. 
	
	Now, let us define a new set of operators 
	\begin{eqnarray}
		J_+ = \hbar ~a_1^\dagger a_2, \; J_- = \hbar ~a_2^\dagger a_1, \; J_z = \dfrac{\hbar}{2}\,(N_1 - N_2). \label{eqn 37}
	\end{eqnarray}
		The algebra for these operators can be obtained as
	\begin{eqnarray}
		[J_z, J_{\pm}] = \pm \hbar ~J_{\pm}, \quad [J_+, J_-] = 2\hbar~ J_z. \label{eqn 38}
	\end{eqnarray}	
	The Casimir operator for this set of operators is obtained as
	\begin{eqnarray}
		J^2 && = J_z^2 + \dfrac{1}{2}\,\lbrace J_+, ~ J_-\rbrace, \nonumber\\
		&& =\dfrac{\hbar^2}{4}\,\dfrac{N}{2}\left(\dfrac{N}{2} +1\right), \label{eqn 39}
	\end{eqnarray}
	where $N= N_1+ N_2$ is the total number operator. 
	But this is the algebra for angular momentum $J$. Thus, the two harmonic oscillator systems are equivalent to a single particle performing orbital motion around a centre. 
	
	The eigenfunctions for the corresponding angular momentum states and that of the harmonic oscillators can be obtained from the mapping as
	\begin{eqnarray}
		\vert j, m\rangle_{AM} = \dfrac{(a_1^\dagger)^{j+m}((a_2^\dagger)^{j-m}}{\sqrt{(j+m)!(j-m)!}}
		\, \vert 0,0\rangle_{SHO}, \label{eqn 40}
	\end{eqnarray}
	where the subscript $AM$ stands for the angular momenta and $SHO$ stands for the simple harmonic oscillators. 
		Here, we have used 
	\begin{eqnarray}
		j = \dfrac{n_1 + n_2}{2}, \quad m = \dfrac{n_1 -n_2}{2}. \label{eqn 41}
	\end{eqnarray}
		The energy of this system of two independent oscillators is given as 
	\begin{eqnarray}
		H &&= H_1 + H_2, \nonumber\\
		&& = \hbar\omega_1 \left(n_1 + \dfrac{1}{2}\right) + \hbar\omega_2 \left(n_2 + \dfrac{1}{2}\right). \label{eqn 42}
	\end{eqnarray}
		It is worth mentioning here that we can obtain the algebra for the fermionic oscillators in the same way. 
	
	Let us define the number operators for the fermionic oscillators as
	\begin{eqnarray}
		\tilde N_1 = f^\dagger_1 f_1, \quad \tilde N_2 =  f^\dagger_2 f_2, \label{eqn 43}
	\end{eqnarray} 
	where $f$ and $f^\dagger$ are the annihilation and creation operators for the fermionic state functions. The algebra of fermionic oscillators for $i =1,2$ is as follows
	\begin{eqnarray}
		\lbrace f^\dagger_i,  f_i \rbrace = 1, \; \lbrace \tilde N_i, f_i \rbrace = -f_i, \;\lbrace \tilde N_i, f^\dagger_i\rbrace = f^\dagger_i, \label{44}
	\end{eqnarray}
	with 
	\begin{eqnarray}
		&& \tilde N_1 \vert \tilde n_1, \tilde n_2\rangle = \tilde n_1\,\vert \tilde n_1, \tilde n_2\rangle, \nonumber\\ 
		&& \tilde N_2 \vert \tilde n_1, \tilde n_2\rangle = \tilde n_2\,\vert \tilde n_1, \tilde n_2\rangle. \label{eqn 45}
	\end{eqnarray}
		Here for the fermionic oscillators
	\begin{eqnarray}
		f \vert 0\rangle = 0, ~ f^\dagger \vert 0\rangle = \vert 1\rangle, ~ f \vert 1\rangle = \vert 0\rangle, ~ f^\dagger \vert 1\rangle = 0. \label{eqn 46}
	\end{eqnarray}
	
	The corresponding angular rotor momentum operators are defined as
	\begin{eqnarray} 
		&& J_+ \equiv \hbar \,f_1^\dagger f_2, ~ J_- \equiv \hbar\, f_2^\dagger f_1, \nonumber\\
		&& J_z \equiv \dfrac{\hbar}{2}\, (f_1^\dagger f_1 - f_2^\dagger f_2) =  \dfrac{\hbar}{2}~ ({\tilde N_1} - {\tilde N_2}). \label{eqn 47}
	\end{eqnarray}
		If we compute the Casimir operator, it turns out to be on the fermionic system as
	\begin{eqnarray}
		J^2 && = J_z^2 + \dfrac{1}{2}\lbrace J_+, ~J_-\rbrace,\nonumber\\
		&& = \hbar^2 ~\dfrac{\tilde N}{2}\left(\dfrac{\tilde N}{2}  + 1\right) - \hbar^2\lbrace{\tilde N_1,~ \tilde N_2\rbrace},\nonumber\\
		&& = \hbar^2 ~\dfrac{\tilde N}{2}\left(\dfrac{\tilde N}{2}  + 1\right) - 2\hbar^2 {\tilde N_1 \tilde N_2}, \label{eqn 48}
	\end{eqnarray}
	where ${\tilde N} = {\tilde N_1} + {\tilde N_2} $ and $[\tilde N_1,~\tilde N_2] = 0.$
	
	Since the eigenvalues of $\tilde N_1$ corresponding to the states $\vert 0\rangle_1$ and $\vert 1\rangle_1$ are 0 and 1 respectively, and similarly for $\tilde N_2$ too. Therefore, the possible values of Casimir operator $J^2$ can either be 0 (for the combined states $\vert 0,0\rangle $ and $\vert 1,1\rangle $) or $3\hbar^2/4$ (for $\vert 0,1\rangle $ and $\vert 1,0\rangle $). The corresponding $j$-values for the combined states, namely, $\vert 0\rangle_1 \otimes \vert 0\rangle_2 \equiv \vert 0,0\rangle, ~\vert 1,1\rangle $ and $\vert 0,1\rangle, ~\vert 1,0\rangle $ can be obtained as follows: 
	\begin{eqnarray}
		&&\tilde N \vert 0,0\rangle, \nonumber\\
		&&= (\tilde N_1 + \tilde N_2)\vert 0\rangle_1 \otimes \vert 0\rangle_2, \nonumber\\
		&&= (\tilde N_1\vert 0\rangle_1)\otimes \vert 0\rangle_2 + \vert 0\rangle_1 \otimes (\tilde N_2\vert 0\rangle_2), \nonumber\\
		&& = (0\vert 0\rangle_1)\otimes \vert 0\rangle_2 + \vert 0\rangle_1 \otimes (0\vert 0\rangle_2), \nonumber\\
		&& = 0 \vert\ 0,0\rangle. \label{eqn 49}
	\end{eqnarray}
	and similarly, 
	\[\tilde N\vert 0,1\rangle = 1 \vert\ 0,1\rangle, ~\tilde N\vert 1,0\rangle = 1 \vert 1,0\rangle, ~\tilde N\vert 1,1\rangle = 2 \vert 1,1\rangle \label{eqn 50}\]
	yields the eigenvalue equations for the Casimir operator $J^2$ to be 0 for the combined states $\vert 0,0\rangle, ~\vert 1,1\rangle$ states while $3\hbar^2/4$ for the states $\vert 0,1\rangle, ~\vert 1,0\rangle$ states. However, from the usual angular momentum algebra, the Casimir operator $J^2$ must have the eigenvalue equations as:
	\[J^2 \vert j,m\rangle = \hbar^2 j(j+1)\vert j, m\rangle. \label{eqn 51}\]
	
	This implies that the equivalent expression for the eigenvalue $j$ is $(0, \hbar/2)$. The values of $m$ corresponding to $j = 0, 1/2$ are $m= 0, \pm 1/2$, i.e.
	\[J_z \vert 0,0\rangle = 0, ~ J_z \vert \dfrac{1}{2}, \pm \dfrac{1}{2}\rangle = \pm \dfrac{\hbar}{2}\vert \dfrac{1}{2}, \pm \dfrac{1}{2}\rangle. \label{eqn 52} \]
	
	The good thing about this algebra is that it applies equally to a system of bosonic and fermionic oscillators. We can show that the algebra for such a collection of bosonic plus fermionic oscillators is similar. For this purpose, let us define a set of new operators such that 
	\begin{eqnarray} 
		&& J_+ \equiv \hbar\,a^\dagger f, ~J_- \equiv \hbar f^\dagger a, \nonumber\\
		&& J_z \equiv \dfrac{\hbar}{2}(a^\dagger a - f^\dagger f) = \dfrac{\hbar}{2} (N -\tilde N). \label{eqn 53}
	\end{eqnarray}	
	It can be shown that the algebra for these mixed operators is 
	\begin{eqnarray}
		[J_z, ~J_\pm]  = \hbar J_\pm, ~[J_+, J_-] = 2\hbar J_z. \label{eqn 54}
	\end{eqnarray}	
	The corresponding Casimir operator can be obtained as follows: 
	\begin{eqnarray}
		J^2 && = J_z^2 + \dfrac{1}{2} \lbrace J_+, ~ J_-\rbrace, \nonumber\\
		&& = \dfrac{\hbar^2}{4} (a^\dagger f - f^\dagger a)^2 + \dfrac{\hbar^2}{2}\,(a^\dagger f f^\dagger a + f^\dagger a a^\dagger f), \nonumber\\
		&& = \dfrac{\hbar^2}{4}(N^2 + \tilde N - 2\tilde N N) + \dfrac{\hbar^2}{2} (N + \tilde N),\nonumber\\
		&& = \dfrac{\hbar^2}{4} \left(\tilde N(\tilde N + 2) + N(N + 2) - \lbrace N, \tilde N\rbrace\right). \label{eqn 55}
	\end{eqnarray}	
	For a fermionic oscillator, the number operator is a projection operator, i.e. ${\tilde N}^2 = \tilde N$. Therefore, there can be only two states, which we have already discussed above, i.e. $\vert 0\rangle $ and $\vert 1\rangle$.
	
	The combined states are $\vert n, 0\rangle$ and $\vert n,1 \rangle $ only. The corresponding eigenvalues of $J^2$ for these states are as follows:
	\[J^2 \vert n, 0\rangle = \dfrac{\hbar^2}{4} n(n+2) \vert n, 0\rangle, ~ J^2\vert n, 1\rangle = \dfrac{\hbar^2}{4} (n^2 +3)\vert n,1\rangle. \label{eqn 56}\]
	
	This is a bit different from the previous angular momentum algebras as here we do not get the usual structure of angular momentum states as $ J^2\vert j, m\rangle = \hbar^2 j(j+1)\vert j, m\rangle.$
	
	However, there is a way to commensurate this discrepancy by redefining the ladder operators regarding the bosonic and fermionic creation and annihilation operators \cite{Mejia}. For this, let us learn a theorem which will be very useful in defining things further. For some arbitrary real number $r$, we can use
	\begin{eqnarray}
		a N^r = (1+N)^r a, ~~ N^r a^\dagger = a^\dagger (1+N)^r. \label{eqn 57}
	\end{eqnarray}
		With this result, we can redefine the creation, annihilation and $J_z$ operator as follows:
	\begin{eqnarray}
		&& J_+ = \hbar a^\dagger (1+N)^{-1/2} f, ~J_- = \hbar f^\dagger (1+N)^{-1/2} a, \nonumber\\
		&& J_z = \dfrac{\hbar}{2}\left(a^\dagger (1+N)^{-1} a (1-\tilde N) -\tilde N\right), \label{eqn 58}
	\end{eqnarray}
	which, with help of the formula mentioned in (\ref{eqn 53}), can be rewritten as
	\begin{eqnarray} 
		&& J_+ =\hbar N^{-1/2} a^\dagger f, ~ J_- = \hbar f^\dagger a N^{-1/2}, \nonumber\\
		&& J_z = \dfrac{\hbar}{2}\, (1-2\tilde N). \label{eqn 59}
	\end{eqnarray}
	
	It is easy to check that the action of these operators on a combined state $\vert n, \tilde n \rangle \equiv \vert n\rangle \otimes \vert \tilde n \rangle$, are obtained as 
	\begin{eqnarray}
		&& J_+ \vert n, \tilde n\rangle = \hbar \sqrt{\tilde n}\vert n+1, \tilde n\rangle; \tilde n =1, \nonumber\\ 
		&& J_- \vert  n, \tilde n\rangle = \hbar \sqrt{1-\tilde n}\vert n+1, \tilde n\rangle; \tilde n = 0 , \nonumber\\
		&& J_z  \vert  n, \tilde n\rangle = \dfrac{\hbar}{2} (1- 2\tilde n) \vert  n, \tilde n\rangle. \label{eqn 60}
	\end{eqnarray}	
	With help of the operators given in (\ref{eqn 59}), we can construct the corresponding Casimir $J^2$ as
	\[ J^2 \equiv J_z^2 + \dfrac{1}{2}\lbrace J_+,~J_-\rbrace =  \dfrac{3\hbar^2}{4} I, \label{eqn 61}\]
	which in turn when operates on a combined state $ \vert \vert n, \tilde n\rangle$ yields the result 
	\[J^2  \vert  n, \tilde n\rangle = \hbar^2 \, \dfrac{1}{2} \left(\dfrac{1}{2} +1 \right) \vert n, \tilde n\rangle, ~\label{eqn 62}\]
	with a fixed eigenvalue $j=1/2$, constrained with the coupling of angular momenta of $j_1 = 0, j_2 = 1/2$ and hence $j = 1/2$ only. 
	
	The states in combined Hilbert space can be obtained as follows 
	\begin{eqnarray}
		\vert n, {\tilde n}\rangle = \dfrac{(a^\dagger)^n}{\sqrt{n!}}\vert 0,~\tilde n\rangle; ~~\tilde n = 0,1.  \label{eqn 63}
	\end{eqnarray}
	
	\section{Physical Realization of Schwinger Oscillators}\label{sec8}
	A physical realization of the Schwinger oscillators can be obtained by taking them as points on a rotating circle of radius equal to the amplitude of the oscillators (provided all of them have the same amplitude). The radius of gyration can be obtained classically as well as quantum mechanically. However, The radius of this can be obtained indirectly by calculating the Schwinger angular momenta of the two oscillators. A formula for the radius of gyration can be obtained in (\ref{eqn 67}) for various angular momentum states. The treatment done above was for the independent bosonic oscillators.

	\subsection{Collection of bosonic oscillators}
	Let us consider a collection of harmonic oscillators oscillating with corresponding frequencies.  As we know from Schwinger's proposition any pair of oscillators taken together is algebraically equivalent to an angular momentum algebra of some rotor. If there are $n$ such oscillators, then there could be $n \choose 2$ pairs of oscillators. Each such pair would have an equivalent angular momentum algebra. These individual angular momenta can couple with each other through Clebsch-Gordan coefficients. Physically, we can realize a collection of harmonic oscillators as $n\choose 2$ rotors. All these equivalent rotors can couple each other in $n\choose 2$ ways. For a collection of bosonic oscillators, any two can be viewed as a single particle moving around some centre. Now, one can ask if an equivalent single particle is performing some circular motion. Then, what is its moment of inertia? As a wise guess, we can take it as a function of the masses of individual oscillators, i.e. $\mu \equiv \mu (m_i, m_j).$ Then the moment of inertia of this equivalent single rotor will be ${\cal I} = \mu r^2$, where $r$ is the radius of the orbit. How shall we determine the moment of inertia? 
	
	There is a way to do this. As we know, for a system of two bosonic oscillators, the total energy is 
	\begin{eqnarray}
		E_v = \dfrac{\hbar}{2}\, (n_1~\omega_1 + n_2~\omega_2 + 2). \label{eqn 64}
	\end{eqnarray}
	Since this energy must be equivalent to that of the rotor, i.e. 
	\[E_r = \dfrac{\hbar^2}{2\cal I}~ j(j+1), \label{eqn 65}\]
	where the relation between oscillator quantum numbers $n_1, n_2$ and $j$ are already discussed above. 
	
	Now, to simplify the calculations, let us consider that all the oscillators are identical and in ground states. Then, with this, the moment of inertia of the equivalent rotor can be calculated as follows 
	\begin{eqnarray}
		&& \dfrac{\hbar^2}{2\cal I}\, 1(1+1) = 2\times \dfrac{\hbar \omega}{2}, \nonumber\\
		\Rightarrow && ~{\cal I} = \dfrac{\hbar}{\omega}. \label{eqn 66}
	\end{eqnarray}
	
	If we take ${\cal I} = \mu k^2$ for the reduced mass $\mu$ and $k$ as the radius of gyration and $\tilde \omega$ as the angular frequency of rotation, then the angular momentum square would be 
	\begin{eqnarray}
		J = {\cal I} \tilde\omega = \dfrac{\hbar\tilde \omega}{\omega}.  \label{eqn 67}
	\end{eqnarray}
	Squaring and taking the ground state of rotation, i.e. $j=1$, we shall have 
	\[\tilde \omega = \sqrt 2 ~\omega. \label{eqn 68}\]
	
	Thus, with the help of the Schwinger proposition, one can extract a lot of information, viz. the equivalent moment of inertia, equivalent radius of gyration and frequency of rotor in terms of the inputs from oscillators. 
	
	We can take three identical bosonic oscillators as well. All are oscillating with fundamental frequency in ground states. Then, the vibrational energy can be counted to be 
	\begin{eqnarray}
		E_v = 3\times \dfrac{3}{2}\hbar\omega =   \dfrac{9}{2}\hbar\omega. \label{eqn 69}
	\end{eqnarray}
	
	Since any two oscillators can be taken to be equivalent to a rotor, one will be left out. If the frequency of the rotor is $\tilde\omega$, then the energy of the rotor plus the oscillator would be
	\begin{eqnarray}
		E_r = \dfrac{\hbar^2}{2\cal I} + \dfrac{\hbar\omega}{2}. \label{eqn 70}
	\end{eqnarray} 
	
	Comparing the two energies, we end up with the same conclusion $\tilde\omega = \sqrt 2 ~\omega$.
	
	This mechanism can be applied to any number of oscillators. And the equivalent rotational frequencies can be calculated subsequently. 
	
	Since the Schwinger formalism does not say anything about the locality of oscillators, whether the collection is local or global, we can safely say that this relation is true locally and globally. It does not matter how far two oscillators are. 
	
	If there is a huge number of oscillators in a thermal bath at some equilibrium temperature $T$, then the partition function can be obtained as
	\begin{eqnarray}
		Z[0] && = {\text Tr}~e^{-\beta H} = \sum_{n=0}^\infty \langle n| e^{-\beta H}| n\rangle, \nonumber\\
		&& 	= \sum_{n=0}^\infty \langle n| e^{-\hbar\beta\omega(n+ 1/2)}| n\rangle, \nonumber\\
		&& 	= e^{-\hbar\beta\omega/2}\sum_{n=0}^\infty e^{-n\hbar\beta\omega} \langle n| n\rangle, \nonumber\\
		&& 	= e^{-\hbar\beta\omega/2}\sum_{n=0}^\infty e^{-n\hbar\beta\omega}, \nonumber\\
		&& 	= \dfrac{e^{-\hbar\beta\omega/2}}{1-  e^{-\hbar\beta\omega}}, \nonumber\\
		-\ln Z[0] && = \dfrac{\hbar\beta\omega}{2} + \ln \left(1- e^{-\hbar\beta\omega}\right). \label{eqn 71}
	\end{eqnarray}

	The formula obtains the corresponding energy 
	\begin{eqnarray}
		\langle E_{vib}\rangle && = -\dfrac{\partial}{\partial\beta}~\ln Z[0],\nonumber\\
		&& = \dfrac{\hbar\omega}{2} + \hbar \omega \left(\dfrac{e^{-\hbar\beta\omega}}{1 - e^{-\hbar\beta\omega}}\right), \nonumber\\
		&& = \hbar\omega\left(\dfrac{1}{2} + \dfrac{e^{-\hbar\beta\omega}}{1- e^{-\hbar\beta\omega/2}}\right). \label{eqn 72}
	\end{eqnarray}
	
	Now, as we know, there are $n\choose 2$ pairs of harmonic oscillators which can be taken equivalent to a rotor. An equivalent rotor exists for any pair of oscillators. For these $n\choose 2$ rotors, there exist corresponding $j$-values. For each pair of harmonic oscillators oscillating in various modes, the average energy of the rotor could be obtained by the relation
	\[\langle E_{rot}\rangle = -\dfrac{\partial}{\partial\beta}~\ln Z_{rot}[0], \label{eqn 73}\]
	where 
	\[Z_{rot}[0] = \sum_{j=0}^{n\choose 2}(2j+1)~e^{-\beta \epsilon_j}, \label{eqn 74}\]
	with
	\[\epsilon_j = \dfrac{\hbar^2}{2\cal I}~ j(j+1), \label{eqn 75}\]
	for j-values obtained from Schwinger formalism. 
	
	However, there is one more way to compute the energy of equivalent rotors by considering the coupling of individual angular momenta to yield the resultant angular momentum $J$ through this formula 
	\[\vec J = \sum_{a,b =1}^n \dfrac{\vec J_{ab}}{2n}, \label{eqn 76}\]
	where the factor of 2n is taken to avoid any overcounting. The formula can obtain the corresponding energy.  
	\[\epsilon_j = \dfrac{\hbar^2}{2\cal I}~ j(j+1). \label{eqn 77}\]
	
	Thus, we can see the collection of bosonic oscillators as a giant rotor rotating with some angular momentum $J$ obtained by the equivalence of energies and Schwinger formalism. 
	
	\subsection{Collection of fermionic oscillators}
	
	If we stick with only two fermion oscillators, then it becomes easy to compute the energy of this system. Let us look into this explicitly. 
		The formula gives the average energy for a pair of fermionic oscillators in a thermal bath at $T$.
	\[\langle E\rangle  = -  \dfrac{\hbar}{2}~(\omega_1 + \omega_2) + \dfrac{\hbar\omega_1}{1+ e^{\beta\hbar\omega_1}} + \dfrac{\hbar\omega_2}{1+ e^{\beta\hbar\omega_2}}. \label{eqn 78}\]
		Comparing with the energy obtained from the formula 
	\[E = \dfrac{\hbar^2}{2\cal I}\, j(j+1). \label{eqn 79}\]
		Comparing the total energy of the pair of oscillators, one can compute the equivalent moment of inertia of the equivalent rotor.

	\section{Discussion}\label{sec9}
	The one-dimensional harmonic oscillator oscillates in real one-dimensional space, which may or may not be linear. A bosonic harmonic oscillator oscillating in its $n^{th}$ mode is equivalent to a collection of n-identical harmonic oscillators oscillating in their fundamental modes. The Hilbert space of an oscillator in its $n^{th}$ mode is  equivalent to the direct/tensor product of individual ground state Hilbert spaces of individual oscillators, i.e.
	\[{\cal H}_n = {\cal H}_1 \otimes {\cal H}_1 \otimes \cdots \equiv ({\cal H}_1) ^{n}. \label{eqn 80}\]	
		In the case of a fermionic oscillator, the space in which oscillation occurs is not spatial but rather something intrinsic. Also, the values of $n_1$ and $n_2$ in the case of fermionic oscillators are not larger than 1 since the fermions consist of only two states, namely $\mid 0\rangle$ and $\mid 1\rangle$. That is why estimating the continuum limit of a fermionic oscillator becomes non-trivial for a given total kinetic energy. Unlike the bosonic case, the energy calculation for the fermionic case using Schwinger formalism is not straightforward because the Pauli exclusion principle and energy due to this have not been incorporated into the calculation, which should have been.

	\section{Summary and Future Works}\label{sec10}
	In this work, we have demonstrated a significant generalization of Schwinger's correspondence between the algebra of harmonic oscillators and angular momentum. Our findings reveal that this correspondence is not limited to bosonic oscillators but extends universally to any two harmonic oscillators, whether they are bosonic-bosonic, fermionic-bosonic, or fermionic-fermionic. This result marks a profound breakthrough, showing that the Schwinger scheme applies far beyond its traditionally understood scope, encompassing a more comprehensive algebraic framework. The implications of this generalization are far more substantial than initially anticipated, opening up new avenues in theoretical and applied physics.

In addition to this, we have rigorously derived the Lagrangian and Hamiltonian for fermionic oscillators from first principles. While the Lagrangian is rooted in the Dirac Lagrangian with its intrinsic Grassmannian nature, we observe a critical modification after quantization. Specifically, while classical Grassmann fields adhere strictly to Grassmann algebra, their corresponding quantum operators need not obey the same algebra. This highlights a fundamental difference in the behavior of classical and quantum systems within this context, offering new insights into the nature of fermionic oscillators.

Furthermore, we have uncovered an intriguing result: fermionic oscillators oscillate at half the rate of their bosonic counterparts within the same time period. While this result may initially seem counterintuitive, it is consistent with the well-known fact that bosonic systems exhibit a double covering of fermionic loops. Thus, a fermion requires two full oscillations to return to its original configuration, akin to completing a M\"obius-like twisted loop to complete a full orbit. This unexpected behavior underscores the distinctive nature of fermionic systems and adds to our understanding of their dynamics.

Looking ahead, our next step is to apply this generalized formalism to practical problems involving extensive collections of diverse oscillators. We aim to verify these results through computational methods and cross-check them with predictions from quantum statistical mechanics. This endeavour promises to provide new tools for analyzing complex systems where various oscillators coexist, potentially yielding important insights for both fundamental research and practical applications.
	
\section {Acknowledgement}
Dheeraj Shukla  is extremely grateful to Ashok Das, Dheeraj Mishra  and Ramlal Awasthi for their valuable discussions and suggestions.

\end{document}